\documentclass[aps,prl,reprint,groupedaddress,twocolumn]{revtex4-1}
\usepackage{graphicx,amsmath,color}
\begin{document}

%\preprint{}

\title{Self-induced polar order of active Brownian particles 
%with hydrodynamic interactions\\
in  a harmonic trap}

%\title{Self-induced order of active Brownian particles with hydrodynamic interactions in an external field}

\author{Marc Hennes}
\author{Katrin Wolff}
%\email[]{katrin.wolff@tu-berlin.de}
\author{Holger Stark}
\affiliation{Institut f\"ur Theoretische Physik, Technische Universit\"at Berlin, Hardenbergstra\ss e 36, 10623 Berlin, Germany}

\date{\today}

\begin{abstract}
Hydrodynamically interacting active particles in an external harmonic potential
form a self-assembled fluid pump at large enough P\'eclet numbers. Here,
we give a quantitative criterion for the formation of the pump and show that 
particle orientations align in the self-induced flow field in surprising 
analogy to ferromagnetic order where the active P\'eclet number plays the role 
of inverse temperature. The particle orientations follow a Boltzmann 
distribution 
%$\Phi(p_z)\sim\exp(A p_z)$
%$\Phi(p_z)\sim\exp(A \cos \theta)$ so mischt man doch Zylinder- mit 
%Kugelkoordinaten, ich bin fur $p_z$ auch im exp}, 
$\Phi(\mathbf{p}) \sim \exp(A p_z)$
where the ordering mean 
field $A$ scales with active P\'eclet number and polar order parameter.
The mean flow field in which the particles' swimming directions align 
corresponds to a regularized stokeslet with strength proportional to swimming 
speed. Analytic mean-field results are compared with results from Brownian 
dynamics simulations with hydrodynamic interactions included and are found 
to capture the self-induced alignment very well.
\end{abstract}

\maketitle

\paragraph{Introduction}

Understanding the non-equilibrium behavior of self-propelled particles is one of
the major challenges at the interface of physics, biology, and also chemical 
engineering~\cite{cates_diffusive_2012,romanczuk_active_2012}. Interacting 
active particles may show exotic phenomena such as swirling 
motion~\cite{dunkel_fluid_2013}, or exhibit dynamic 
clustering~\cite{theurkauff_dynamic_2012,palacci_living_2013} and 
motility-induced phase separation
\cite{tailleur_statistical_2008,fily_athermal_2012,redner_structure_2013,
buttinoni_dynamical_2013,stenhammar_continuum_2013}. Their collective motion
drives macroscopic fluid flow as in 
bioconvection~\cite{pedley_hydrodynamic_1992} 
or vortex formation~\cite{ordemann_pattern_2003}. Hydrodynamic interactions
between microswimmers crucially determine their collective patterns
\cite{aditi_simha_hydrodynamic_2002,ishikawa_coherent_2008,
saintillan_instabilities_2008,baskaran_statistical_2009,
evans_orientational_2011,
%zottl_nonlinear_2012,
molina_hydrodynamic_2013,
alarcon_spontaneous_2013, zottl_hydrodynamics_2013},
while external fluid flow leads to aggregation~\cite{torney_transport_2007},
trapping~\cite{khurana_reduced_2011}, and nonlinear swimming 
dynamics~\cite{zottl_nonlinear_2012}.

In order to understand the collective dynamics of active particles and their 
steady-state distributions, they are often mapped onto passive systems 
%moving 
that move
% oder ganz weglassen?} 
in effective potentials~\cite{tailleur_sedimentation_2009,enculescu_active_2011, 
wolff_sedimentation_2013}. However, for interacting particles there is no 
general route for identifying an equilibrium 
counterpart~\cite{tailleur_statistical_2008,cates_diffusive_2012,
stenhammar_continuum_2013}.

The system we investigate here is composed of self-propelled or active Brownian
particles whose swimming directions undergo rotational diffusion in a harmonic 
trap. Bacteria or both active and passive colloids confined in optical 
traps have attracted 
experimental~\cite{chattopadhyay_swimming_2006,lutz_surmounting_2006,
lincoln_reconfigurable_2007} as well as
theoretical~\cite{schweitzer_complex_1998,rex_dynamical_2008,
tailleur_sedimentation_2009,nash_run-and-tumble_2010,pototsky_active_2012} 
interest. Passive colloids are operated in non-equilibrium by switching the 
trapping force~\cite{rex_dynamical_2008} while active particles are 
intrinsically out of 
equilibrium~\cite{tailleur_sedimentation_2009,nash_run-and-tumble_2010,
pototsky_active_2012}. Run-and-tumble particles in lattice Boltzmann 
simulations develop a pump state which breaks the rotational symmetry of the 
harmonic trap and cause a macroscopic fluid 
flow~\cite{nash_run-and-tumble_2010}. Here, we demonstrate similar behavior for 
active Brownian particles which interact by hydrodynamic flow 
fields. However, more importantly we explain the emerging orientational order 
of particles by mapping the self-induced alignment of swimmers in a harmonic 
potential onto an equilibrium system which exhibits ferromagnetic order. 

To this end we first establish a quantitative criterion for the formation of the 
pump and then introduce a mean-field description for the fully formed pump 
state. The mean-field system shows a striking analogy to the Weiss molecular 
field in ferromagnetism and reproduces our Brownian dynamic simulations.

The system also bears some similarity to the vortex formation in \emph{Daphnia} 
populations~\cite{ordemann_pattern_2003} caused by irradiation with light where 
the apparent attraction towards the center of the light spot has also been modeled
by a harmonic potential~\cite{vollmer_vortex_2006}. The crucial difference, however, is that 
\emph{Daphnia} swim towards the light by phototaxis whereas in the system 
discussed here the harmonic potential exerts a body force on the swimmers. One 
focus of this work is thus on the self-induced polar order of active particles 
and how it is mapped on a passive system with very different underlying physics.

\paragraph{The model}

We consider a dilute suspension of $N$ self-propelled particles with 
constant propulsion speed $v_0$ whose leading hydrodynamic interactions are 
modeled by (far-field) mobility tensors $\boldsymbol\mu_{ij}$. Particles are 
spherical with an internal orientation vector $\mathbf{p}_i$, as realized, for 
example, in active colloids \cite{palacci_living_2013,buttinoni_dynamical_2013} 
and they swim with velocity $v_0 \mathbf{p}_i$. The Langevin equations of motion 
for the position $\mathbf{r}_i$ and orientation $\mathbf{p}_i$ of particle $i$ 
then are
\begin{equation*}
\dot{\mathbf{r}}_i=\mathbf{v}_i,\quad \dot{\mathbf{p}}_i=\boldsymbol\omega_i\times\mathbf{p}_i
\end{equation*}
with
%{\color[rgb]{1,0,0}{
\begin{eqnarray*}
 \mathbf{v}_i  & = & v_0\mathbf{p}_i + 
\sum_{j=1}^N \boldsymbol\mu_{ij}^{tt}\mathbf{F}_j + \sum_{j=1}^{2N}
  \mathbf{H}_{ij}\boldsymbol\xi_j
  +
 \sum_{j\neq i}^N \mathbf{u}_{\mathrm{SS},j}(\mathbf{r}_{ij})
\nonumber \\
 \boldsymbol\omega_i & = & \sum_{j=1}^N \boldsymbol\mu_{ij}^{rt}\mathbf{F}_j +
 \sum_{j=1}^{2N}\mathbf{H}_{(i+N)j}\boldsymbol\xi_j +
 \sum_{j\neq i}^N \boldsymbol\omega_{\mathrm{SS},j}(\mathbf{r}_{ij}).
\nonumber
\end{eqnarray*}
%}}
%
%\begin{equation*}
% \mathbf{v}_i = v_0\mathbf{p}_i + \sum_{j}
% \left( 
%\boldsymbol\mu_{ij}^{tt}\mathbf{F}_j +
%  \mathbf{H}^t_{ij}\boldsymbol\xi_j
% \right) +
% \sum_{j\neq i} \mathbf{u}_{\mathrm{SS},j}(\mathbf{r}_i-\mathbf{r}_j)
%\end{equation*}
%and
%\begin{equation*}
%\boldsymbol\omega_i =  \sum_j \left(\boldsymbol\mu_{ij}^{rt}\mathbf{F}_j +
% \mathbf{H}^r_{ij}\boldsymbol\eta_j\right) +
% \sum_{j\neq i} \boldsymbol\omega_{\mathrm{SS},j}(\mathbf{r}_i-\mathbf{r}_j).
%\end{equation*}
%
The particle velocity $\mathbf{v}_i$ thus consists of the self-propulsion term 
$v_0 \mathbf{p}_i$ along the particle's orientation vector, the contribution 
from the harmonic trapping forces acting on all particles,
$\mathbf{F}_j=-k_\mathrm{trap}\mathbf{r}_j$, a stresslet velocity term 
$\mathbf{u}_\mathrm{SS}$ caused by the active swimming of all other particles
with $\mathbf{r}_{ij}=\mathbf{r}_i - \mathbf{r}_j$, and the thermal noise 
term $\sum_{j=1}^{2N}\mathbf{H}_{ij}\boldsymbol\xi_j$. Here, 
$\boldsymbol{\xi}_j$ is 
a $3$-component vector and contains time-uncorrelated Gaussian variables with 
zero mean and unit variance for translational or rotational noise. The 
noise components are coupled to each other by the $3\times 3$ amplitude matrices 
$\mathbf{H}_{ij}$, which are determined by the fluctuation-dissipation theorem. 
Further details are found in the supplemental material. The swimmers interact 
hydrodynamically via the second and fourth term in $\mathbf{v}_i$. We consider 
hydrodynamic interactions up to  second order in $1/r$, with $r$ the swimmer 
distance, and use for the translational mobility $\boldsymbol\mu_{ij}^{tt}$ the 
Oseen tensor. The velocity stresslet 
$\mathbf{u}_{\mathrm{SS},j}(\mathbf{r})=\beta v_0 \frac{3a^2}{4 r^2} 
[-3(\mathbf{p}_j \cdot \hat{\mathbf{r}})^2 +1] \,\hat{\mathbf{r}}$, with 
$\hat{\mathbf{r}} = \mathbf{r} /r$ and particle radius $a$, determines whether a 
particle is an extensile (pusher, $\beta<0$), contractile (puller, $\beta>0$) or 
neutral ($\beta=0$) swimmer. Unless stated otherwise, we set $\beta=0$ and use 
the neutral swimmer as default.

Angular velocity likewise consists of a deterministic part due to the vorticity 
caused by the trapping force, the vorticity due to the swimmers' flow dipoles 
$\boldsymbol\omega_\mathrm{SS}=1/2(\nabla\times\mathbf{u}_\mathrm{SS})$, and 
thermal rotational noise $\mathbf{H}_{(i+N)j}\boldsymbol\xi_j$. The mobility 
tensor $\boldsymbol\mu_{ij}^{rt}$ couples translational to rotational motion and 
in leading order of $1/r^2$ is given in the supplemental material. We simulate 
the collective dynamics of active particles using Brownian dynamics simulations 
with hydrodynamic interactions following an extended form of the algorithm by 
Ermak and McCammon~\cite{ermak_brownian_1978}, where the self-propulsion and 
swimming terms are included.

\begin{figure}
\includegraphics[width=\linewidth]{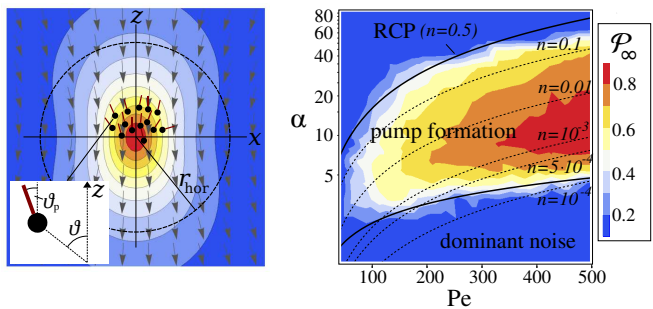}%
\caption{\label{fig:pumpformation}(Color online.) \emph{(Left)} Sketch of 
the fluid pump. Active particles concentrate in the upper half sphere and align 
their orientations along the vertical. The resulting flow field is illustrated 
by a regularized stokeslet. The dashed black circle with radius 
$r_\mathrm{hor}$ is the particles' horizon. \emph{(Right)} State diagram of 
pump formation in active ($\mathrm{Pe}$) versus trapping ($\alpha$) P\'{e}clet 
numbers. The mean orientation of the particles, $\cal{P}_{\infty}$, is 
color-coded. The lower solid curve, $\alpha = 4\sqrt{\mathrm{Pe}/(\pi N)}$, 
separates the region of pump formation from the region of dominant 
rotational noise. Along the dashed lines density $n$ is constant. At very 
high densities the active particles are close-packed and cannot form a pump 
(upper solid curve). }
\end{figure}

\paragraph{Pump formation}

In steady state, non-interacting active particles accumulate at or near the 
surface of a sphere where active swimming and trapping force cancel each 
other. The particles's horizon has the characteristic radius 
$r_\mathrm{hor} = a \mathrm{Pe} / \alpha$~\cite{tailleur_sedimentation_2009},
where $a$ is the particle radius, $\mathrm{Pe}= v_0 a / D$ the active P\'eclet 
number, and $\alpha = k_\mathrm{trap} a^2 / (k_BT)$
the trapping P\'eclet number with translational diffusion coefficient $D= k_BT / 
(6\pi \eta a)$.
%$\alpha = k_\mathrm{trap}a / (6\pi\eta D)$ 
The spherical symmetry, however, is broken when hydrodynamic interactions are 
included. At sufficiently large $\mathrm{Pe}$, particles assemble into a pump 
in a more tightly packed region, where they align their swimming directions 
and produce a macroscopic fluid flow~\cite{nash_run-and-tumble_2010}. The 
situation is sketched in Fig.~\ref{fig:pumpformation} (\emph{left}) and 
accompanying videos can be found in the supplemental material.

Figure~\ref{fig:pumpformation} (\emph{right}) gives the state diagram for 
pump formation. A simple criterion for pump formation against thermal noise 
(lower solid line in the diagram) can be derived from a comparison of time 
scales. In simulations, particles first accumulate mainly at the characteristic 
radius $r_\mathrm{hor}$ of the non-interacting system without generating any 
macroscopic fluid flow. So translational advection but also translational noise 
are not important at this point, however, their orientations can freely diffuse.
Therefore, the relevant criterion for particles to assemble into a pump is the 
following. When hydrodynamic interactions between particles are strong enough to 
overcome rotational diffusion, they rotate particles' swimming directions 
towards spontaneously formed denser particle regions. Thus if the time scale of 
rotational diffusion, $T_\mathrm{diff}=2\pi^2/D_R$~\footnote{For thermal 
diffusion of spherical particles, the rotational ($D_R$) and translational ($D$) 
diffusion coefficients are connected by $D_R=3D/(4 a^2)$}, is larger than the 
rotation time due to the flow field's vorticity 
$T_\mathrm{HI}=2\pi/\omega_\mathrm{HI}$, particles will be rotated towards dense 
spots and then, by swimming towards them, further enhance those density hot 
spots. Estimating the net vorticity disturbance on a particle in the otherwise 
spherically symmetric initial state to be $\omega_\mathrm{HI}=k_\mathrm{trap} 
r_\mathrm{hor} / (8\pi \eta r_\mathrm{nn}^2)$ with the nearest-neighbor distance 
$r_\mathrm{nn}\approx4 r_\mathrm{hor} / \sqrt{N}$, we find the pump formation 
criterion $\alpha\geq4\sqrt{\mathrm{Pe}/(\pi N)}$. Details of the calculations 
are given in the supplemental material. 

This rough estimate reproduces the onset of pump formation quite well [see 
lower solid line in Fig.~\ref{fig:pumpformation}\ (\emph{right})]. To indicate 
the alignment of particle orientations in the steady state, we use the polar 
order parameter
$\mathcal{P}_\infty=\lim_{t\rightarrow\infty}\left|\sum_{i=1}^N \mathbf{p}_i(t)\right|/N$
\cite{evans_orientational_2011}. For very dense systems, particles become 
close-packed and no pump is formed as the upper solid curve at constant 
density $n=0.5$ in Fig.~\ref{fig:pumpformation} (\emph{right}) shows. At such 
high densities, however, the assumption of far-field interactions also no longer 
holds. In the following we concentrate on dilute systems and, unless stated 
otherwise, follow the path of the dashed line with mean density $n=10^{-3}$. To 
realize a constant density for different $\mathrm{Pe}$, we fix the particle 
number $N=100$ and keep the effective trap volume constant by varying 
the trapping P\'{e}clet number $\alpha$ alongside the active P\'eclet number 
$\mathrm{Pe}$ to ensure 
$r_\mathrm{hor}\propto\alpha/\mathrm{Pe}=\mathit{const.}$.

\paragraph{Mean-field theory for self-induced order}

In order to gain analytical insight, we reduce the Smoluchowski equation for the 
$N$-particle distribution function to a mean-field equation for the one-particle 
distribution $\psi(\mathbf{r},\mathbf{p})$. The collective dynamics due to 
hydrodynamic interactions is taken into account by a mean flow field 
$\mathbf{u}(\mathbf{r})$ generated by all the particles. It is independent 
of time, once the pump has formed. The Smoluchowski equation governing the 
effective one-particle dynamics is
\begin{equation}
 \partial_t \psi(\mathbf{r},\mathbf{p}) = - \nabla \cdot \mathbf{J}_T - \mathcal{R} \cdot \mathbf{J}_R
 \quad
 \textrm{with}
  \quad\mathcal{R}=\mathbf{p}\times\nabla_\mathbf{p} .
 \label{eq:smoluchowski}
\end{equation}
We use the translational flux $\mathbf{J}_T = [v_0 \mathbf{p}+\mu_t 
\mathbf{F}_\mathrm{ext}(\mathbf{r}) + \mathbf{u}(\mathbf{r}) ]  
\psi(\mathbf{r},\mathbf{p})$ and the rotational flux
$\mathbf{J}_R = [ (\nabla\times\mathbf{u}(\mathbf{r}))/2 - D_R\mathcal{R} ] \psi(\mathbf{r},\mathbf{p})$.
Translational diffusion is neglected in $\mathbf{J}_T$ because of P\'{e}clet 
numbers $\mathrm{Pe}\gtrsim100$, whereas rotational diffusion still needs to be 
included. As we assume the flow field $\mathbf{u}$ to be independent of time, 
Eq.~(\ref{eq:smoluchowski}) only describes the dynamics close to a fully formed 
pump state and we will just attempt to determine the steady state distribution 
$\psi(\mathbf{r},\mathbf{p})$.

\begin{figure}
\includegraphics[width=0.8 \linewidth]{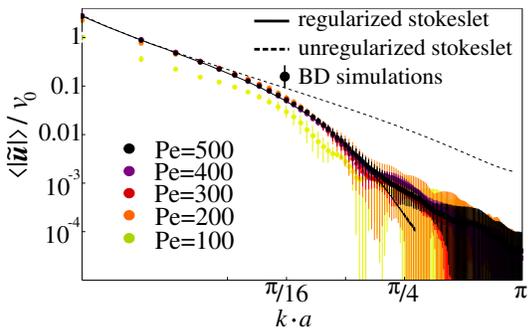}%
\caption{\label{fig:FTfig}(Color online.) 
Directional average for the absolute value of the Fourier transform of fluid 
velocity, $\langle \left| \tilde{\mathbf{u}}(\mathbf{k})  \right| \rangle$,
plotted versus wave number $k$. Results obtained from the Brownian dynamics 
simulations (colored dots) are compared with a regularized stokeslet (solid 
line) and a stokeslet without regularization (dashed line).
}
\end{figure}

Strictly speaking, the collective flow field follows self-consistently from the 
distribution $\psi(\mathbf{r},\mathbf{p})$ by integrating over all particle 
contributions, 
$\mathbf{u}(\mathbf{r})=\int\boldsymbol{\mu}(\mathbf{r}-\mathbf{r}')
%\cdot
\mathbf{F}_\mathrm{ext}(\mathbf{r'})\psi(\mathbf{r'},\mathbf{p'})d\mathbf{r}'d\mathbf{p'}$.
Here, however, we first determine the flow field from the full $N$-particle
simulations, argue that it is well represented by a regularized stokeslet, and
only check \emph{a posteriori} that our resulting mean-field density gives rise
to the same flow field.

To quantify the flow field $\mathbf{u}(\mathbf{r})$, Fig.\ \ref{fig:FTfig} shows 
the directional average for the absolute value of its Fourier transform 
$\tilde{\mathbf{u}}(\mathbf{k})$. The average goes over all directions of wave 
vector $\mathbf{k}$ keeping wave number $k$ fixed. Before taking the Fourier 
transform, the flow field has been averaged over 100 uncorrelated simulation 
snapshots in the steady state to suppress fluctuations. We find that in the 
parameter range, where the pump is fully formed, the strength of the flow field 
$\mathbf{u}$ is simply proportional to the particle swimming speed $v_0$. As 
Fig.~\ref{fig:FTfig} demonstrates, for active P\'{e}clet number 
$\mathrm{Pe}=100$ the pump has not yet fully formed but for $\mathrm{Pe}=200$ to 
500 all data fall on a single master curve when rescaled by $v_0$. They agree 
very well with the flow field of a regularized 
stokeslet~\cite{cortez_method_2005},
\begin{equation}
 \mathbf{u}_\mathrm{reg}(\mathbf{r}) = 
    -\frac{v_0\epsilon}{2(r^2+\epsilon^2)^{3/2}}
   \left[ \mathbf{1}(r^2 + 2\epsilon^2)+\mathbf{r}\otimes\mathbf{r} \right]
   \mathbf{e}_z,
\label{eq.reg}
\end{equation}
also for the regions \emph{inside} the pump (see Fig.~\ref{fig:FTfig}, solid 
line). Here, the flow field deviates from an unregularized stokeslet (dashed 
line) and only in the far field (small $k$) does the flow field become a 
conventional stokeslet. The regularization parameter $\epsilon$ is used as a 
fit parameter but it turns out that it coincides with the radius of the region 
populated by active particles and thus can be interpreted as the pump radius. 
For large wave numbers $k$, the simulated flow field starts to differ from 
$\mathbf{u}_\mathrm{reg}$ as fluctuations on the scale of several particle radii 
become visible. 

\begin{figure}
\includegraphics[width=0.78\linewidth]{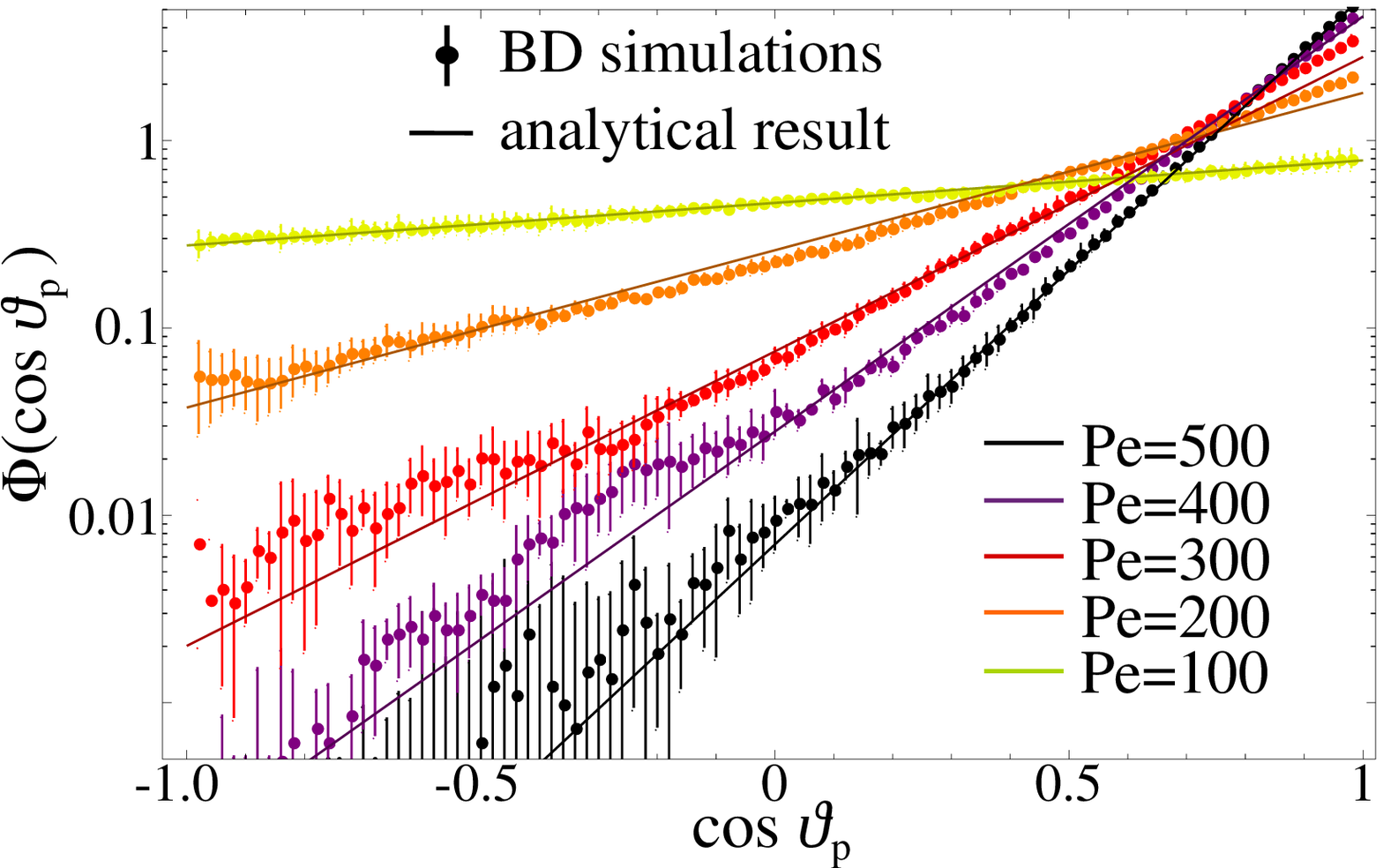}%

\includegraphics[width=0.79\linewidth]{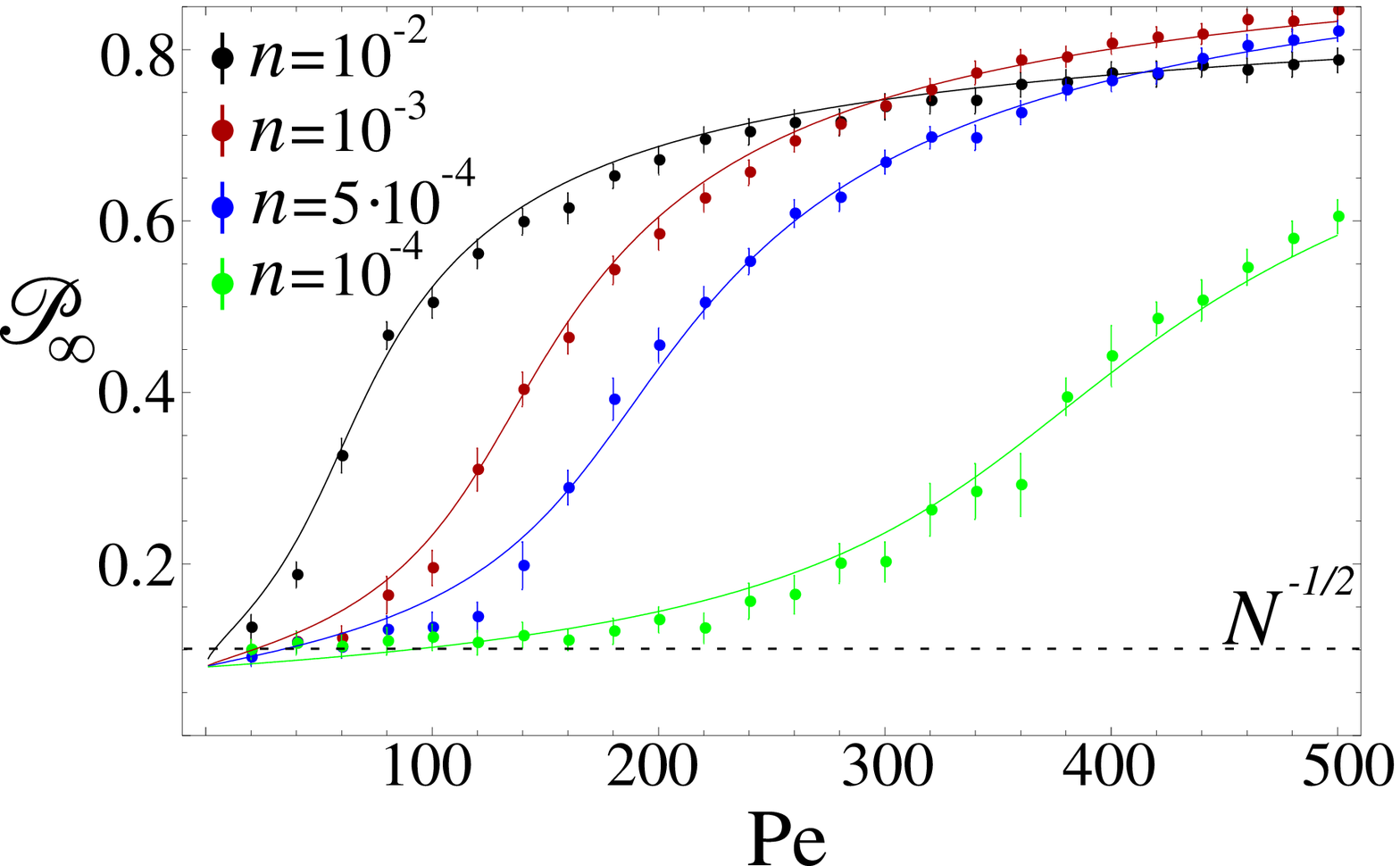}%
\caption{\label{fig:orientations}(Color online.) (\emph{Top}) 
Orientational distribution functions for different values of particle activity 
$\mathrm{Pe}$ at fixed average density $n=10^{-3}$. The simulation results 
(dots) are very well reproduced by $\Phi(\cos\theta_p)\sim \exp(A\cos\theta_p)$ 
(solid lines), where $A$ can be derived analytically. (\emph{Bottom}) Particle 
alignment $\mathcal{P}_\infty$ for increasing activity $\mathrm{Pe}$ at 
different densities. The simulation results (dots) again agree very well with 
the analytic result 
$(\mathcal{P}_\infty-N^{-1/2})/\mathcal{P}_\infty^\mathrm{max} 
= \mathcal{L}(3\,(\mathrm{Pe}\,\mathcal{P}_\infty / 
\mathrm{Pe}_\mathrm{c})^\gamma)$
(solid lines).
}
\end{figure}

We now make an ansatz for the particle distribution $\psi(\mathbf{r},\mathbf{p})$
and assume that particle orientations always point radially outward, 
$\psi(\mathbf{r},\mathbf{p})=\Phi(\mathbf{p})\,f(r)\,\delta(\cos(\theta)-\cos(\theta_p))\delta(\varphi-\varphi_p)$.
Here $\theta$ and $\varphi$ are the spherical coordinate angles of position and 
$\theta_p$ and $\varphi_p$ those of orientation. The polar angles $\theta$  and 
$\theta_p$ are both measured against the main axis of the pump [see also 
the sketch in Fig.~\ref{fig:pumpformation} (\emph{left})]. The assumption of 
parallel orientation and position vector is motivated by the system without 
hydrodynamic interactions at high P\'eclet numbers. Here, particles quickly 
swim to the horizon where the trapping force cancels their self-propulsion. 
With interactions included, the assumption of parallel orientation and 
position vector still holds approximately (see supplemental material). 
Integrating Eq.~(\ref{eq:smoluchowski}) over particle position using the 
ansatz for $\psi(\mathbf{r},\mathbf{p})$, we arrive at the equation for the 
orientational distribution function,
\begin{equation}
 \partial_t\Phi(\mathbf{p}) = -\mathcal{R}  \cdot   [ \langle \boldsymbol \omega 
\rangle(\mathbf{p}) - D_R \mathcal{R} ] \Phi(\mathbf{p}),
% \partial_t\Phi(\mathbf{p}) = -\mathcal{R}\cdot\langle\boldsymbol\omega\rangle(\mathbf{p})
%  + D_R\mathcal{R}^2 \Phi(\mathbf{p})
  \label{eq:oridistribution}
\end{equation}
%\begin{equation}
%% \partial_t\Phi(\mathbf{p}) = -\mathcal{R}  \cdot   [ \langle \boldsymbol \omega \rangle(\mathbf{p}) - D_R \mathcal{R} ] \Phi(\mathbf{p})
% \partial_t\Phi(\mathbf{p}) = -\mathcal{R}\cdot\langle\boldsymbol\omega\rangle(\mathbf{p})
%  + D_R\mathcal{R}^2 \Phi(\mathbf{p})
%  \label{eq:oridistribution}
%\end{equation}
where the mean vorticity of the collective flow field is determined for the 
regularized stokeslet of Eq.\ (\ref{eq.reg}):
$\langle\boldsymbol\omega\rangle(\mathbf{p}) = - D_R A 
\,\, \sin(\theta_p) \, \mathbf{e}_{\varphi_p}$ 
with $A= \mathrm{Pe} \int_0^\infty 
\frac{\epsilon\,a(5\epsilon^2+2r^2)}{3(\epsilon^2+r^2)^{5/2}}r^3  f(r)\mathrm{d}r$.
Eq.~(\ref{eq:oridistribution}) can then be solved in steady state and the 
orientational distribution function becomes 
$\Phi(\mathbf{p})=\mathrm{e}^{A\cos(\theta_p)}/\mathcal{N}$,
where $\mathcal{N}$ is a normalization factor. Figure\ \ref{fig:orientations} 
(\emph{top}) shows $\Phi(\mathbf{p})$ for different active P\'{e}clet numbers 
in very good agreement with Brownian dynamics simulations. The radial 
distribution $f(r)$ necessary to determine the constant $A$ for the analytic 
result has been extracted from simulations.

All particles thus create a mean flow field in which single swimmers align, in 
analogy to Weiss theory for ferromagnetism using a molecular field. Here the 
regularized stokeslet or more precisely the field strength $A$ takes the role of 
the molecular magnetic field and mean polar order $\cal{P}_{\infty}$ that 
of magnetization. This analogy can be made more explicit by looking at the 
overall alignment $\mathcal{P}_\infty = \int_0^{2\pi}\int_{-1}^1 
\mathbf{p}\,\Phi(\mathbf{p}) \mathrm{d}\varphi_p\mathrm{d}\cos(\theta_p)$,
which evaluates to $\mathcal{P}_\infty = \mathcal{L}(A) = \coth(A) - 1/A$ with 
the Langevin function $\mathcal{L}(A)$ as encountered in the classical theory 
of magnetism \cite{neil_w._ashcroft_solid_1976}. Importantly, the field strength 
$A$ depends in turn on the alignment $\mathcal{P}_\infty$ and we find in our 
simulations $A\propto(\mathrm{Pe}\,\mathcal{P}_\infty)^\gamma$ with an exponent 
$\gamma$ close to 1 for low densities and decreasing for higher densities (see 
supplemental material). So, we indeed have a formal analogy to the Weiss 
molecular field generalized to an exponent $\gamma\leq 1$. The active 
P\'eclet number $\mathrm{Pe}$ takes the role of inverse temperature and a 
critical $\mathrm{Pe}$ can be determined.

Fig.~\ref{fig:orientations} (\emph{bottom}) shows alignment curves for various
values of density $n$. Due to the finite number of particles $N$, the curves are 
shifted from 0 to $\mathcal{P}_\infty = N^{-1/2}$ for no alignment. Similarly, 
total alignment $\mathcal{P}_\infty = 1$ cannot be attained because the 
shape of the pump and the excluded volume of the particles prohibit completely 
parallel particle orientations. The latter is taken into account via a 
geometric parameter $\mathcal{P}_\infty^\mathrm{max}$. The implicit equation 
for $\mathcal{P}_\infty$ becomes 
$(\mathcal{P}_\infty-N^{-1/2})/\mathcal{P}_\infty^\mathrm{max} 
= \mathcal{L}(3\,(\mathrm{Pe}\,\mathcal{P}_\infty / 
\mathrm{Pe}_\mathrm{c})^\gamma)$, which fits the data very well when 
numerically solved for $\mathcal{P}_\infty$ with the critical P\'eclet number 
$\mathrm{Pe}_\mathrm{c}$ and the geometric parameter 
$\mathcal{P}_\infty^\mathrm{max}$ as fitting parameters. The exponent 
$\gamma=1.0$ for $=10^{-4}$ but decreases for higher densities and is only 
$\gamma=0.5$ for $n=10^{-2}$, which is likely due to excluded volume effects. 
The critical P\'eclet numbers for the different densities are found to agree 
roughly with the values derived from the time scale argument in 
Fig.~\ref{fig:pumpformation}.

\begin{figure}
\includegraphics[width=0.8\linewidth]{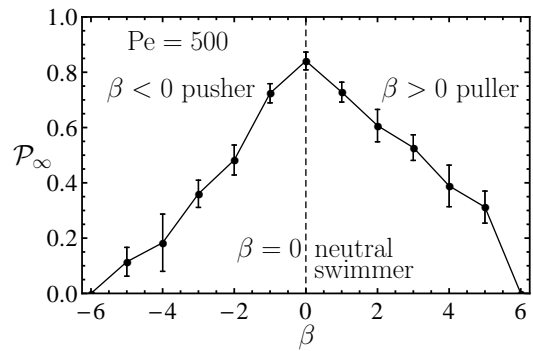}%
\caption{\label{fig:beta} Influence of swimmer dipole fields on the pump 
formation at activity $\mathrm{Pe}=500$ and volume fraction $n=10^{-3}$. 
Mean polar order $\mathcal{P}_\infty$ plotted versus swimmer type $\beta$.
}
\label{fig4}

\end{figure}

We also investigated the effect of the velocity dipoles ($\beta \ne 0$) due to 
the particles' swimming and found those flow stresslets to hinder alignment
(Fig.\ \ref{fig4}). This is in agreement with previous studies on suspensions 
of active particles~\cite{evans_orientational_2011,alarcon_spontaneous_2013}. 
Neutral swimmers ($\beta=0$) show the strongest ordering meaning that the 
alignment of swimmers is mediated solely by the stokeslets due to the trapping 
force, as assumed in our calculations. Our studies also show an asymmetry 
between pushers and pullers with alignment decreasing faster for pushers 
($\beta<0$) than pullers $\beta>0$. Again this agrees with previous 
studies~\cite{evans_orientational_2011,alarcon_spontaneous_2013} which 
explained this asymmetry by the observation that head-to-head orientations 
(which would contribute to $\mathcal{P}_\infty=0$) are stabilized for pushers 
due to a stagnation point in front of the swimmer in the swimming frame.

\paragraph{Conclusions}

We investigated the pump formation of active Brownian particles in a harmonic 
trap with hydrodynamic interactions included. This is an example of a 
non-equilibrium system whose properties are in striking analogy to a well-known 
equilibrium system. Specifically, we showed that the self-induced alignment 
of particles mediated by the flow field they create is in formal agreement 
with spontaneous magnetization in ferromagnetic materials treated on the 
mean-field level. Here, particle orientations follow a Boltzmann distribution 
in an aligning mean flow field which is created by the active particles. A 
critical P\'{e}clet number, where pump formation sets in, can be determined in 
analogy to the (inverse) critical temperature of ferromagnets. The mean flow 
field corresponds to a single regularized stokeslet. Its strength scales
linearly with the particle swimming speed and the regularization parameter 
$\epsilon$ gives the pump radius.

% The alignment of particles is the result of the interplay of the (radially 
% symmetric) external trapping force causing strong hydrodynamic interactions
% between the particles and the particles' activity which drives the system out of 
% equilibrium. The flow dipoles created by the swimming particles, however, do not
% contribute to alignment and instead produce fluctuations in the flow field which
% counteract alignment.

Understanding the non-equilibrium of active particles is one of the challenging 
questions in statistical physics right now. This Letter presents an intriguing 
example of a non-equilibrium system of active particles which can be mapped onto 
a classical equilibrium system. With recent advances in colloid physics it 
would be very interesting to experimentally realize the fluid pump formed 
by self-induced polar order and validate our predictions.

\paragraph{Acknowledgments}
\begin{acknowledgments}
We thank Andreas Z\"ottl for helpful discussions and gratefully acknowledge
financial support from the Deutsche Forschungsgemeinschaft through the research 
training group GRK1558.
\end{acknowledgments}

\bibliography{pumpformation}

\end{document}